# Dosimetric investigation of $^{103}$Pd permanent breast seed implant brachytherapy based on Monte Carlo calculations

Stephen G. Deering[a] (MSc), Michelle Hilts[b,c] (PhD), Deidre Batchelar[b,c] (PhD), Juanita Crook[d] (MD), and Rowan M. Thomson[a,1] (PhD)

[a]Carleton Laboratory for Radiotherapy Physics, Department of Physics, Carleton University, Ottawa, Canada K1S 5B6

[b]Department of Medical Physics, BC Cancer-Kelowna, Kelowna, British Columbia, Canada

[c]Department of Physics, IK Barber School of Arts and Sciences, University of British Columbia - Okanagan, Kelowna, British Columbia, Canada

[d]Department of Radiation Oncology, BC Cancer –Kelowna, Kelowna, British Columbia, Canada

[1]Corresponding author at: Department of Physics, Carleton University, 1125 Colonel By Drive, Ottawa ON K1S 5B6 Canada, Canada; E-mail: rowan.thomson@carleton.ca

Conflict of interest: None.

Funding: The present study was supported by an Early Researcher Award from Ontario's Ministry of Research and Innovation, the Canada Research Chairs program, the Natural Sciences and Engineering Research Council of Canada, and Compute Canada.

Data sharing: Research data are not available at this time.

Acknowledgements: The authors thank Daniel Morton for discussions.

Running title: MC dose evaluations for breast brachytherapy

## Abstract

**Purpose:** Permanent breast seed implant (PBSI) using $^{103}$Pd is emerging as an effective adjuvant radiation technique for early-stage breast cancer. However, clinical dose evaluations follow the water-based TG-43 approach with its considerable approximations. Towards clinical adoption of advanced TG-186 model-based dose evaluations, this study presents a comprehensive investigation for PBSI considering both target and normal tissue doses.

**Methods and Materials:** Dose calculations are performed with the free open-source Monte Carlo (MC) code, egs_brachy, using 2 types of virtual patient models: **TG43sim** (simulated TG-43 conditions: all water with no interseed attenuation) and **MCref** (heterogeneous tissue modelling from patient CT, interseed attenuation, seeds at implant angle) for 35 patients. Sensitivity of dose metrics to seed orientation and the threshold for glandular/adipose tissue segmentation are assessed.

**Results:** In the target volume, $D_{90}$ is 14.1±5.8% lower with **MCref** than with **TG43sim**, on average. Conversely, normal tissue doses are generally higher with **MCref** than with **TG43sim**, e.g., by 22±13% for skin $D_{1cm}{}^2$, 82±7% for ribs $D_{max}$, and 71±23% for heart $D_{1cm}{}^3$. Discrepancies between **MCref** and **TG43sim** doses vary over the patient cohort, as well as with the tissue and metric considered. Doses are sensitive to the glandular/adipose tissue segmentation threshold with differences of a few percent in target $D_{90}$. Skin doses are particularly sensitive to seed orientation, with average difference of 4% (maximum 28%) in $D_{1cm}{}^2$ for seeds modelled vertically (egs_brachy default) compared with those aligned with implant angle.

**Conclusions:** TG-43 dose evaluations generally underestimate doses to critical normal organs/tissues while overestimating target doses. There is considerable variation in **MCref** and **TG43sim** on a patient-by-patient basis, suggesting that clinical adoption of patient-specific MC dose calculations is motivated. The **MCref** framework presented herein provides a consistent modelling approach for clinical implementation of advanced TG-186 dose calculations.

# Introduction

Permanent breast seed implant (PBSI) brachytherapy using low-dose-rate (LDR) $^{103}$Pd seeds is an attractive, single-day technique for adjuvant radiation in early-stage breast cancer (1). PBSI has been shown to be an effective treatment option with high rates of patient satisfaction and very good cosmesis (2;3).

Clinical dosimetric results are determined using the formalism of Task Group (TG) 43 of the American Association of Physicists in Medicine (AAPM) (4). This formalism neglects important dosimetric effects of tissue composition and inhomogeneities, influences of the brachytherapy applicators or seeds, as well as the finite extent of the patient. Breast brachytherapy in general has been identified as a treatment site where neglecting these factors can have a significant impact on dose calculation accuracy, with LDR breast brachytherapy being particularly sensitive to tissue composition and interseed attenuation (5). Preliminary studies indicate that TG-43 calculations can overestimate doses to the PTV by as much as 35% while skin doses are underestimated (6;7;8;9). To achieve accurate dosimetry, advanced model-based dose calculation algorithms (MBDCAs) are being developed, and their clinical adoption is recommended by the joint TG-186 of the AAPM–European Society for Therapeutic Radiotherapy and Oncology (ESTRO)–Australasian Brachytherapy Group (10).

Early studies applying Monte Carlo (MC) to PBSI dose calculations have investigated issues related to the development of the virtual patient models needed for MC simulations, including evaluation of different algorithms to mitigate artifacts in post-implant CT images (7) and subsequent assignment of tissue elemental compositions to voxels (6;7;11). In addition, interseed attenuation has been considered, and patient specific MC doses have been compared to those from TG-43 calculations (6;7;8). These works suggest patient-specific MC models are required for accurate dose evaluation, and highlight the importance of the adoption of a well-defined and consistent MC approach as clinics move towards advanced MBDCAs, but further research is needed.

With the aim of facilitating clinical adoption of advanced MBDCAs for PBSI, herein we present a dosimetric evaluation for a new patient cohort using a recently-developed and released, open-source MC code, egs_brachy (12). Dose metrics are presented for the target, skin, as well as other normal tissues - to our knowledge, the first comprehensive presentation of patient-specific PBSI MC doses in the scientific literature. As MC dose calculations are inherently sensitive to patient modelling assumptions (7), we present a MC approach that follows TG-186 recommendations (10) and builds on the published literature (6;7;11). Results are compared to calculations using TG-43 assumptions. Additionally, we assess sensitivity to the threshold for segmenting adipose and glandular tissues in the breast, and seed orientation.

## Methods and Materials

### Clinical data

Anonymized post-implant CT images (Day 0; GE Lightspeed helical scan at 120 kV, 250 mA; resolution: 2 mm axial and 0.977 to 1.25 mm in-plane) from 35 PBSI patients treated consecutively at BC Cancer-Kelowna between July 2012 and February 2016 are retrospectively analyzed in this REB approved study. Patient characteristics are described elsewhere (3). Each patient had an evaluative target volume (ETV, seroma plus 5mm) contoured (13). Additionally, normal breast, ribs, skin (2 mm, internal from body), ipsilateral lung, heart, and chest wall were contoured to evaluate dose to organs at risk (OAR). Central coordinates of 52 to 118 TheraSeed 200 (Theragenics, Buford, USA) $^{103}$Pd seeds and their air kerma strengths (range 2.3-2.8 U) were recorded along with the angle of implant (range 10º-70º) (3). For all patients, the prescription dose was 90 Gy ($D_{100}$).

**Virtual patient and source models**

Virtual patient models must be developed for MC dose evaluations. These models are based on CT images: the CT calibration curve is used to convert Hounsfield units to mass density for each voxel (*Supplementary Material Table 1*, *Figure 1*) and a tissue assignment scheme is then used to assign mass elemental compositions (10;14) to different tissues (*Supplementary Material Table 1*) according to voxel mass density, with reference to the relevant contour (Fig 1a).

Here we develop virtual patient models from post-implant CT images where the presence of implanted brachytherapy sources results in attenuation artifacts (7;15) that, if not accounted for, can lead to errors when applying the tissue assignment scheme and thus corresponding large dose errors. To ameliorate this, we apply a simple threshold replacement algorithm for metallic artifact reduction (7); note that the CT scanner does not have on-board metallic artifact reduction. In simple threshold replacement, the mass densities for relevant voxels near seeds are compared to a threshold value and if the threshold is exceeded their mass density is replaced with a representative density. Here, mass densities for all voxels within a cylinder (radius = 0.5cm, extending ±2 slices) centered on the seed are compared to threshold of 1.16 g/cm$^3$ (corresponding to the threshold for assignment of calcification, Fig 1a). Voxels that exceed this value have their mass density replaced with the mean patient ETV density, 0.917 g/cm$^3$. Similarly, this mass density replacement is applied to any dark streak artifacts inside the breast using a threshold of 0.800 g/cm$^3$, well below the density of normal breast tissues.

Following metallic artifact reduction, two full-tissue tissue assignment schemes are applied, differing only in the mass density boundary used for segmentation of adipose and glandular tissues, $\rho_{A/G}$. For the reference MC approach, **MCref**, $\rho_{A/G}$ is determined for each patient by fitting a histogram of breast voxel mass densities to two Gaussian distributions representing adipose and glandular tissues and selecting the halfway point between the densities corresponding to the distribution peaks. Resulting patient-specific $\rho_{A/G}$ range from 0.9179 to 0.9700 g cm$^{-3}$, with a mean value of 0.9476 g cm$^{-3}$. To assess calculation sensitivity to $\rho_{A/G}$, the patient cohort mean $\rho_{A/G}$ of 0.9476 g cm$^{-3}$ is applied to all data sets in the second tissue assignment scheme to generate virtual patient models with a fixed tissue boundary, **MCfixed**. As well as the full-tissue models, a homogeneous water (uniform mass density 0.998 g/cm$^3$) virtual patient model with identical voxel boundaries is developed for **TG43sim** simulations.

Additionally, we model three seed orientations to investigate the sensitivity to seed angle (Fig 1c). The most realistic model, with seeds oriented along the patient-specific angle of implant, is used as the standard. This is compared to seeds modelled at an angle of 45º within the transverse plane (adjusted according to left/right side, **MCref-45**) and seeds along the z-axis (**MCref-z** superior-inferior). **MCref-45** serves to approximate the implant angle based on the geometry of the implant procedure, relevant in case implant angle is unknown for other centers/researchers. **MCref-z** represents the default alignment (seed axis aligned vertically, z-direction) in egs_brachy.

**MC simulations**

MC simulations are carried out with the EGSnrc (16) (v2016) application egs_brachy (version 2016.09.01) (12;17). Detailed, previously-benchmarked (12) geometric models of TheraSeed 200 sources are superimposed on the virtual patient model described above to carry out the following MC simulations:

- **MCref** (reference scenario): full-tissue patient model with patient-specific $\rho_{A/G}$ (mass-density threshold for adipose/glandular tissue segmentation); seeds at implant angle, interseed attenuation included
- **TG43sim** (simulation of TG-43 conditions): all-water patient model; seeds at implant angle with no interseed attenuation
- **MCfixed** (assess sensitivity to $\rho_{A/G}$): full-tissue patient model with cohort-average ("fixed") $\rho_{A/G}$ = 0.9476 g/cm$^3$; seeds at implant angle, interseed attenuation included

- **MCref-45** (assess sensitivity to seed in-plane orientation): full-tissue patient model with patient specific $\rho_{A/G}$; seeds at 45°, interseed attenuation included
- **MCref-z** (assess sensitivity to seed orientation): full-tissue patient model with patient-specific $\rho_{A/G}$; seeds along z-axis, interseed attenuation included

Initial photon energies are generated using the $^{103}$Pd spectrum from TG-43 (4). Photon transport is modelled to 1 keV, scoring collision kerma to the local voxel medium (using mass-energy absorption coefficients pre-calculated with EGSnrc application *g*) to approximate dose ($D_{m,m}$, dose-to-medium (10)). MC simulation parameters are consistent with earlier work benchmarking egs_brachy (12;17) (*Supplementary Material Table 3*).

Simulations involve $10^9$ histories, resulting in sub-1% statistical uncertainties on voxel doses in the ETV and the skin (primary OAR); heart $D_{1cm3}$ voxels have sub-1.5% uncertainties (for the 27/35 patients with heart $D_{1cm3}$ greater than 1 Gy). Each simulation requires 8 to 12 computational hours (on a decade-old computer cluster comprised of CPUs of varying speeds); however, uncertainties of order 2% which are sufficient for clinical applications would require less than a minute per patient using a multicore processor (2 GB of RAM needed).

Dose and volume metrics are extracted from dose volume histograms for the target and organs at risk. Dose metrics include $D_{90}$ (minimum dose in the hottest 90% of the ETV), $D_{max}$ (highest dose received by a single voxel within the contours for the skin, ribs, chest, heart), $D_{1cm^2}$ (highest dose received by a contiguous 1 cm$^2$ area in the skin), and $D_{1cm^3}$ (highest dose received by a contiguous 1 cm$^3$ volume in the breast or heart). Volume metrics, $V_x$, represent the fraction of a specified region receiving *x* percentage of the prescribed dose, and include $V_{100}$, $V_{150}$, $V_{200}$ in the ETV, $V_{150}$ for the breast, and $V_5$ for the skin.

Dose metrics from **MCref** are compared with those extracted from the 4 alternate models (Alt) by means of the percent difference computed for each patient for that metric, *M*, as follows:

$$\%\Delta = \frac{M_{MCref} - M_{Alt}}{M_{MCref}} \times 100\%$$

## Results

**Dose distributions**

Patient-specific **MCref** dose distributions demonstrate considerable discrepancies in comparison with **TG43sim** doses (Fig. 2a-c). Discrepancies may be observed throughout patient tissues/volumes, and there are abrupt discontinuities in doses at the interfaces of different tissues (due to their different radiological properties, mass-energy absorption coefficients, arising from different elemental compositions). For the ETV, **MCref** doses are lower than **TG43sim** by up to 40%, with a few exceptions where doses to glandular tissue are higher (e.g., subset of voxels in ETVs shown in Fig.2a,c in yellow-orange) due to the interplay of mass-energy absorption coefficients and fluence differences compared with homogeneous water (**TG43sim**). In contrast, doses to the skin are higher for **MCref** by up to a factor of 2 in comparison to **TG43sim**. Similarly, **TG43sim** typically underestimates doses to other normal tissues e.g., ribs, chest wall.

**Target (ETV) dose metrics**

Voxel doses are generally overestimated over the whole ETV for **TG43sim** compared with **MCref**, reflected in the shapes of dose-volume histograms (DVHs; Fig. 2d-f). Differential DVHs differ for **MCref** and **TG43sim** for the same patient as well as between different patients (Fig. 2(e-g)), with the **MCref** curves typically peaked at a lower dose compared to **TG43sim** for each patient. Across the cohort, ETV dose metrics are consistently overestimated with **TG43sim** compared with **MCref**. Figure 3a shows $D_{90}$ on a patient-by-patient basis over the whole cohort with **TG43sim** consistently higher than **MCref** by 14%, and mean $D_{90}$ is 91.7 ± 19.0 Gy (**MCref**) compared with 105.1 ± 24.4 Gy (**TG43sim**) (Table 1). Correspondingly, target coverage is overestimated by **TG43sim**, as evidenced by volume metrics in Table 1 (for the whole cohort as well as example patients) and Fig. 4.

**Normal tissues**

There is considerable variation in the skin doses between patients, e.g., with $D_{1cm^2}$ varying from a few Gy (min 6.6 Gy (**MCref**), 2.2 Gy (**TG43sim**) for patient #2) to > 100 Gy (max 177 Gy (**MCref**), 160 Gy (**TG43sim**) for patient #30) – Fig. 3b. However, TG43 underestimates $D_{1cm^2}$ skin doses in comparison with **MCref** consistently across the cohort by average 22% ± 13% (Table 1). Maximum skin dose, the highest dose received by a single skin voxel in a patient, is generally underestimated by **TG43sim**, but with some exceptions (figure 4). $D_{max}$ for skin varies dramatically between patients ranging from 7.5 Gy (patient #2) to 892 Gy (patient #8) for **MCref**.

In general, **TG43sim** underestimates doses to normal tissues (ribs, chest wall, heart, lung, breast) – table 1 and figure 4. Large discrepancies are observed for rib and chest wall, for which single voxel maximum doses across the cohort increase by an average of 82.3% (ribs) and 29.6% (chest wall) for **MCref** compared to **TG43sim**. Likewise, the mean $V_{100}$ increases by 84% in the ribs and 43% for chest wall.

Heart doses are generally low, with no heart tissue voxel in any patient receiving more than 50% (45 Gy) of the prescription dose and very few patients (6/35) with non-zero values for heart $V_{10}$. Likewise, mean heart dose evaluated with **TG43sim** is low (0.01 to 0.45 Gy) and the mean heart dose increases by 12 to 93% with **MCref**. Patients with implants in their left breast consistently receive a higher mean heart dose (evaluated with **TG43sim or MCref)**.

Doses to the lung are also very small. Extracted $V_5$ and $V_{20}$ metrics see significant increases (45% and 5.9% on average, respectively) in the transition from **TG43sim** to **MCref** patient models. The corresponding absolute mean **MCref** $V_5$ and $V_{20}$ values are very low, at only 7% and 0.8% respectively.

$V_{150}$ for the entire treated breast is overestimated with **TG43sim** by up to 53% compared with **MCref** (trend similar to those observed for target metrics).

**Sensitivity to virtual patient and source modelling**

Over the patient cohort, changing the mass density used for the adipose/glandular tissue segmentation boundary ($\rho_{A/G}$) from a value specific to each patient (**MCref**) to the cohort-mean $\rho_{A/G} = 0.9476$ g/cm$^3$ used for **MCfixed**, the mean absolute difference in ETV $D_{90}$ values is 1.5%, with few patients having differences > 3% (figure 5). The general trend in target $D_{90}$ variation between **MCref** and **MCfixed** may be related to the change in the relative fraction of tissue assigned to adipose or glandular tissues between the different models: $D_{90}$ decreases with more adipose and increases with more glandular tissue (results not shown). The trends are similar for skin, however, the magnitude of differences in skin metrics between **MCref** and **MCfixed** are generally smaller.

Seeds are modelled at the fiducial angle (of insertion) in the transverse plane for **MCref**, and differences are larger for seeds aligned axially (z) direction (**MCref-z**) compared with 45° within the transverse plane (**MCref-45**) – Fig. 5. Differences in ETV $D_{90}$ for **MCref-z** relative to **MCref** are greater than 2% in over half the cohort, with a maximum %Δ of 8%. Seed orientation can have a considerable effect on skin peak dose and $D_{1cm^2}$ as high skin dose areas may be due to only one or a few seeds in close proximity to the skin, with different seed orientations resulting in even greater proximity. Photons are not emitted isotropically from the Theragenics model 200 seeds (which are not spherically symmetric), and different seed orientations can result in appreciable differences in skin dose metrics. Skin dose metrics for seeds at the 45° angle (**MCref-45**) are in closer agreement with **MCref** than **MCref-z**. The average difference between skin $D_{1cm^2}$ values between **MCref** and **MCref-z** is 4%, with one patient differing by 28%. Rib and chest wall tissues are also often located close enough to seed positions for changes in seed orientation to impact dose. In rib and chest wall tissues, the presence of cortical bone (with its high mass-energy absorption coefficient and density) means that even relatively small changes in photon fluence and spectra can result in considerable dose differences, especially comparing peak dose (in a single voxel) Comparing **MCref-z** and **MCref**, chest wall and rib peak doses differ by up to 47%; **MCref** and **MCref-45** peak doses differ by up to 24%.

## Discussion

It has long been recognized that TG-43 based dose calculations have limited the accuracy of brachytherapy dose determinations. Current efforts to develop MBDCAs for brachytherapy seek to provide a practical, consistent and accurate means of calculating and evaluating dose. As we move towards this goal, it is necessary to compare current techniques to the new in order to understand how prescriptions and OAR dose-limits will need to change to better reflect what is actually being delivered.

This retrospective MC study demonstrates large differences in dose distributions between the clinical, water-based TG-43 approach (**TG43sim**) and our reference patient-specific model-based TG-186 MC (**MCref**) dose calculations for all patients. **TG43sim** consistently overestimates dose in the tumour region, with differences in target $D_{90}$ values routinely exceeding 10 Gy (figure 3, table 1). In healthy tissues, **TG43sim** underestimates dose, with skin doses often misrepresented by 20% or more; heart and lung metrics are tens of percent lower in **TG43sim**; **TG43sim** rib peak doses differ from **MCref** models by greater than a factor of three. These differences between **TG43sim** and **MCref** dose metrics also have a large inter-patient variability, preventing the possible use of these percentage differences as correction factors for existing **TG43sim** methods. This motivates the clinical adoption of MBDCAs with corresponding patient-specific models, such as the MC calculations carried out herein, to accurately account for the effects of non-water tissues and interseed attenuation in realistic patient anatomies and treatment geometries.

A few other studies have investigated MBDCAs for PBSI (7;8;18), but modelling (virtual patient models, seed orientation), organ definition (target as seroma plus practice-specific margin; skin thickness), and dose evaluation (dose calculation algorithm) approaches are different, which means that doses calculated will inherently be different (10). Although this weakens the ability to make direct comparisons between PBSI dose distributions determined in various studies, overall trends for the target and organs at risk may be considered. However, as this is the first PBSI work to present metrics for normal tissues other than skin, it is not possible to compare these results directly to other cohorts.

Within the target, all investigators found that TG43 overestimates $D_{90}$, in agreement with the current study. Miksys et al (7) reported for 4 patients treated at The Ottawa Hospital Cancer Centre (TOHCC; same size ETV) that a **TG43sim** model overestimates $D_{90}$ by 14.6% on average relative to patient-specific MC models, in concurrence with the 14.1% difference observed herein (table 1). Afsharpour et al (8) considered 28 patients treated at Sunnybrook Hospital, reporting results for the target defined as the seroma plus an additional 1 cm margin (ETV herein is seroma + 5 mm), observing a TG43 overestimation of 19.7% relative to MC calculations of $D_{90}$. This larger decrease (compared to that observed herein or by Miksys et al (7)) is due to the larger volume of breast considered for the target and, typically, larger amount of adipose tissue included for which mass-energy absorption coefficients are considerably lower than for glandular tissue or water (7).

Even when using simpler correction methods, TG43 is found to overestimate target doses. For a 140 patient cohort treated at Sunnybrook, Mashouf et al examined using the inhomogeneity correction factor (ICF) method to account for inhomogeneities in TG-43 dose calculations for PBSI (18). They reported that the ICF-correction resulted in a 6% reduction in $D_{90}$ values for the larger target (seroma + 1 cm) relative to uncorrected TG43 (18). While the changes in dose created by the use of an ICF factor agree with the general dosimetric trends observed in patient-specific MC simulation, differences between ICF and TG43 doses are smaller than the differences between MC and TG43 doses, suggesting that the approximate ICF calculations may be underestimating the effects of non-water tissues, patient extent, and seeds.

Skin is the normal tissue most likely to develop complications from PBSI treatments, and as such accurate skin dosimetry is of particular importance. Maximum and $D_{1cm}^2$ skin doses vary considerably on a patient-to-patient basis due to patient anatomy (e.g., breast size) and implant location (target-skin proximity), but $D_{1cm}^2$ is considered a more reliable assessment of patient skin dose because maximum skin dose can be affected by one anomalously high dose voxel (13). Across our whole cohort, $D_{1cm}^2$ is consistently higher for **MCref** than for **TG43sim** (mean 22%), in overall accord with the 18% discrepancy observed for TOHCC patients (7) and trends for the Sunnybrook cohort (6).

MBDCAs are inherently sensitive to patient and treatment-specific modelling assumptions (7;10). Previous work demonstrated sensitivity to tissue elemental composition and complexity in tissue assignment, focusing on target and skin metrics (7). In complement, the current study contributes to advancing the understanding of sensitivity of doses to adipose/glandular tissue segmentation threshold ($\rho_{A/G}$) and seed orientation. The current work is the first (to our knowledge) to assess sensitivity of doses to the adipose/glandular tissue segmentation threshold ($\rho_{A/G}$), with sub-1% discrepancies observed between individualized (**MCref**) and fixed (**MCfixed**) thresholds for most patients, however there are some outliers with differences in $D_{90}$ near 5% (figure 5). As the density and proportion of breast adipose and glandular tissues varies considerably from patient-to-patient, it may be preferable for MBDCAs to use patient-specific $\rho_{A/G}$ for derivation of virtual patient models where possible. CT images were used herein for delineation and identification of tissues, including the segmentation of glandular and adipose tisuses. Factors such as the machine's settings (kVp) and image-processing algorithms, image resolution, breast density, may affect the images and hence the development virtual patient models. Future work may investigate the sensitivity of dose calculations to these parameters.

The current study is also the first, to our knowledge, to consider realistic seed orientations and assess sensitivity of dose-volume metrics to seed orientation. Differences in $D_{90}$ of greater than 2% are seen in a large portion of the cohort when comparing realistic (**MCref**) and axially-modelled (**MCref-z**) seed orientations, with even larger differences observed for skin, rib, and chest wall metrics (figure 5b). Differences are less substantial when seeds are modelled at 45° in the transverse plane (compared with **MCref**). These results motivate modelling seeds at the fiducial implant angle, to most closely approximate seeds' orientation during treatment.

Our current results may be used for quantitative evaluation of the PBSI treatments. Considering **TG43sim** doses, ETV $D_{90}$ > 90 Gy was achieved in 25 of 35 cases (71%), compared to only 20/35 (57%) for **MCref** doses, with 3 cases having **MCref** $D_{90}$ of 60 Gy (patients #7, 27, and 29). This highlights that full clinical adoption of MBDCA dose evaluations may necessitate prescription revision.

For the skin, 43% of patients received $D_{1cm^2}$ > 90 Gy with some over 200 Gy using **MCref**, but these values are 22% lower on average using **TG43**, indicating that current dose limits will also need to be revised going forward with MBDCAs. Rib doses are also underestimated by **TG43**, with the volume receiving > 90 Gy increasing by a factor of 6 with **MCref**. However, adult cortical bone is not considered to be a relatively radiosensitive organ, and previous studies investigating brachytherapy treatments with $^{125}$I sources implanted less than 0.5 cm from the ribs showed no osseous toxicity in a long-term retrospective analysis (19). Heart and lung doses are very low in both **TG43** and **MCref** models.

Mean heart $V_{50}$ and $V_{10}$ values are zero, and mean heart **MCref** doses below 0.60 Gy for all patients, while **MCref** patient lung V5 and V20 values are all < 10%. However, when heart $D_{1cm^3}$ metrics determined using MCref are extracted, some patients have values above 10 Gy. Although this is still well below the guidelines set to prevent pericarditis by the 'Quantitative Analyses of Normal Tissue Effects in the Clinic' (QUANTEC) study (20), it may be of clinical interest because the rate of long-term, post-radiation major coronary events has been linked to heart dose with no apparent lower threshold (21). To prevent radiation induced pneumonitis, QUANTEC sets a limit of mean lung dose at 7 Gy, a much higher value than that received by all BC Cancer PBSI patients (as indicated by the low values of **MCref** patient lung V5 and V20 values).

While PBSI treatments aim to minimize doses delivered to skin and other healthy tissues, there is currently little in the way of complication dose thresholds within the PBSI literature. The only demonstrated association is for skin dose (22), thus, to help position PBSI as a standard of care for early stage breast cancer, a move towards establishing quantitative, accurate, and repeatable dose thresholds is needed. For accurate dose calculations to be widely adopted, a multi-institution retrospective study will be carried out that translates the TG-186 model-based dose calculation (MC simulation) methodology used in the current study to different clinical centers. This will enable comparisons of MC and TG43 doses across different cancer centers (with varying practices/methodologies in relation to seed strength, loading pattern, delivery/implant technique, and so on), and future prospective studies, e.g. to assess possible revision of prescription doses, choice of metrics with TG-186 treatment planning, and more. Treatment outcomes will then be considered in conjunction with accurate dose evaluation. Clinical experience indicates that the current delivered dose to the ETV is extremely effective. Thus, a change in calculation methodology to include patient specific mass-density corrections for adipose/glandular tissue segmentation and corrections for seed angle may show that we are delivering significantly less, but there is no need to increase our prescription to achieve what we thought was required. With regards to normal tissue calculations, we should be cognizant of the fact that we are actually delivering higher doses than we appreciated, especially to skin. This will allow clarification of dose thresholds for toxicity and refine our efforts to minimize complications.

## Conclusions

The present work demonstrates that traditional TG43-based dose evaluation results in systematic inaccuracies in PBSI doses for both target and normal tissues, compared with patient-specific MC, across the cohort considered. Overall, the TG43 approach overestimates doses to the target, and consistently underestimates normal tissue doses (skin, ribs, chest wall, breast, heart, and lung). These discrepancies are considerable, with greater than 10% discrepancies observed for the target dose metrics ($D_{90}$), and greater than 20% $D_{1cm}^2$ for skin and even larger for other normal tissues. This research motivates further work with other cohorts comparing full-tissue patient MC simulations with TG43 calculations for PBSI, towards clinical adoption of TG186 MBDCA, and outcomes studies to connect MC doses and treatment outcomes. Furthermore, this work provides impetus for widespread clinical adoption of advanced MBDCA to avoid systematic errors incurred with the TG43 approach. The presented **MCref** framework implemented with the freely-available open source egs_brachy code, may be useful for clinical adoption of state-of-the-art dose evaluations.

# Figures

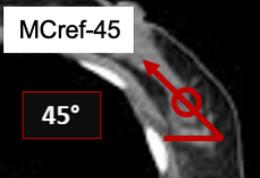

**Figure 1:** Development of simulation geometries: tissue assignment schemes for virtual patient models (**MCref**, **MCfixed**, **TG43sim**) with density boundaries indicated (left) and seed orientations (right) showing an example patient-specific implant angle (55º here) used for **MCref**, **MCfixed**, **TG43sim**; **MCref-45** (45° in transverse plane); **MCref-z** (along z-axis, superior-inferior).

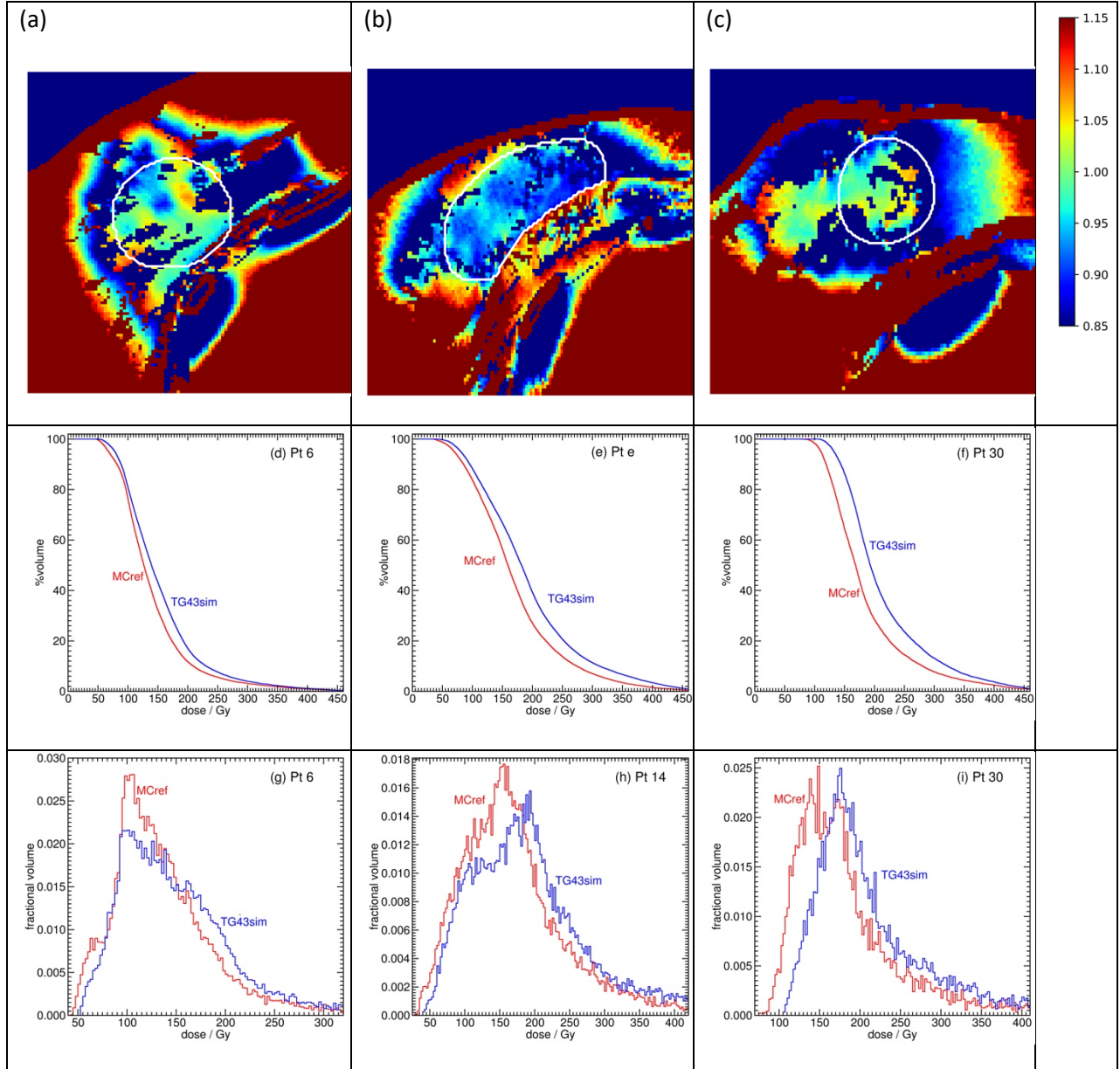

**Figure 2:** Examples of 3 patients (#6, 14, 30) of (a-c) voxel-by-voxel ratios of MCref /TG43 in 1 transverse layer (ETV contour in white); (d-f) dose-volume histogram (DVH); and (g-i) differential DVH in the ETV.

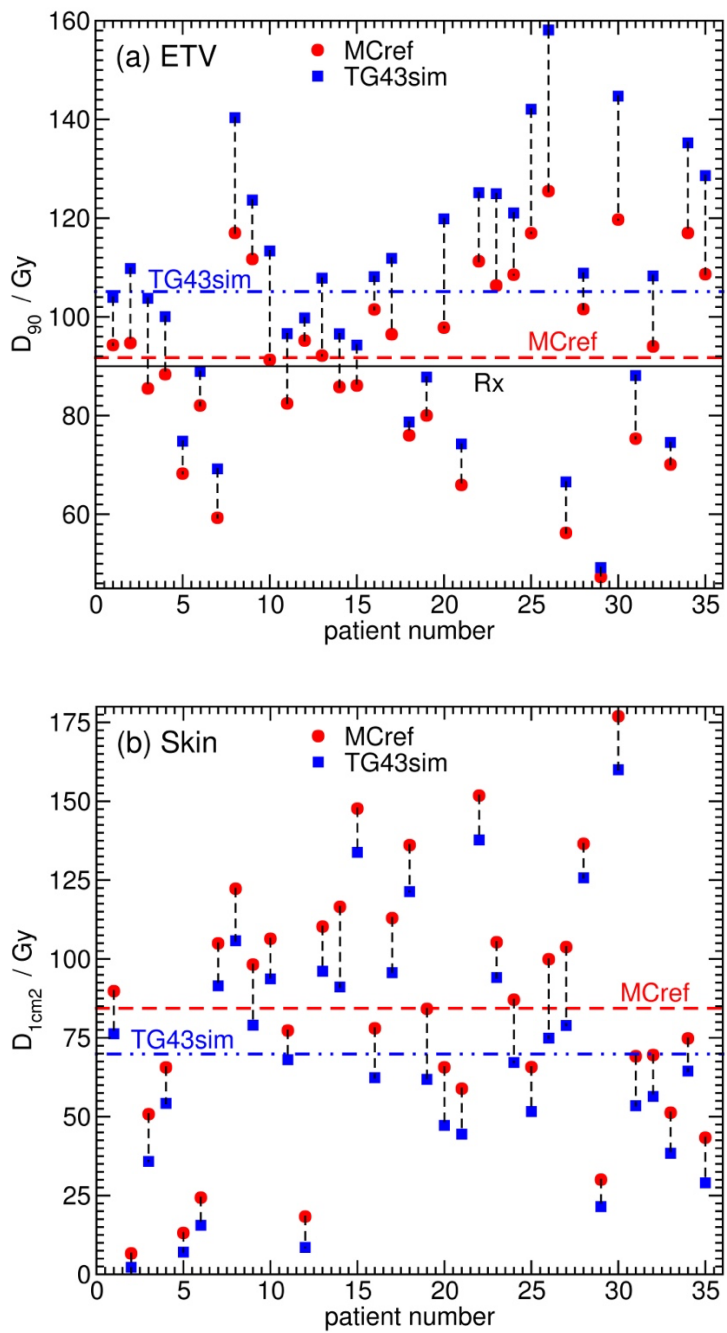

**Figure 3:** Values of (a) ETV (target) $D_{90}$ and (b) skin $D_{1cm^2}$ versus patient number for **TG43** (blue squares) and **MCref** (red circles); horizontal lines indicate mean cohort values (and 90 Gy prescription dose – panel a).

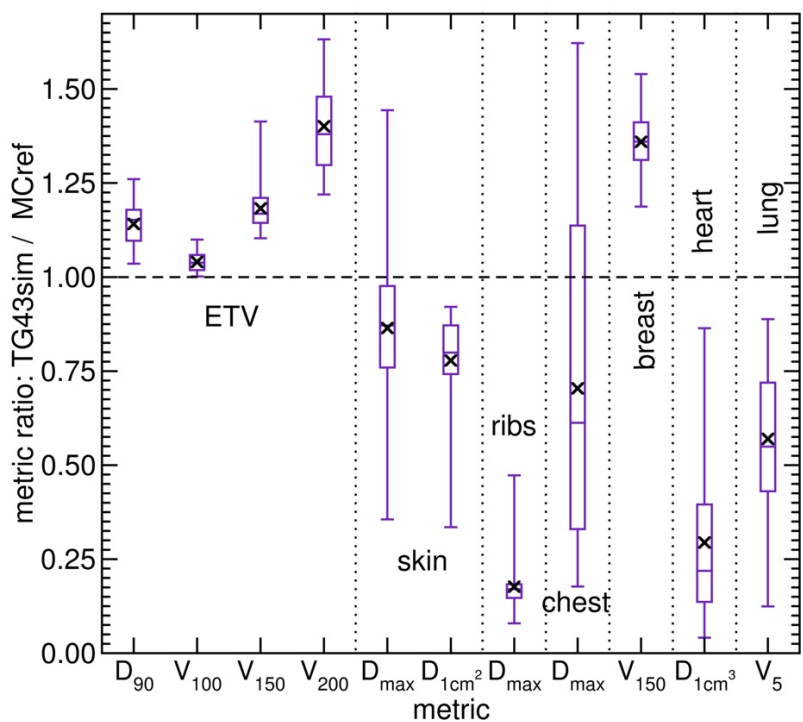

**Figure 4:** Dose metric ratios for **TG43sim** relative to **MCref** for the target (ETV) and normal tissues (skin, ribs, chest, breast, heart and lung) represented by box and whiskers: each box extends from the lower to upper quartile values of the data, with a line at the median, whiskers show the full range of data, and "x" indicates the mean.

21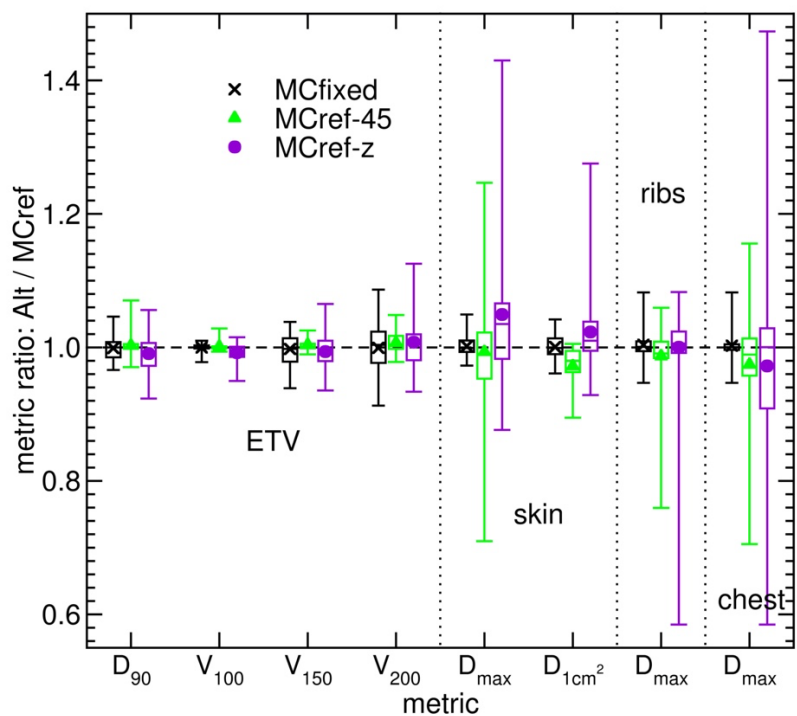

**Figure 5:** Sensitivity to MC modelling: dose metric ratios for alternative MC models (MCfixed, MCref-45, MCref-z) relative to **MCref** for the target (ETV) and some normal tissues (skin, ribs, chest); box and whiskers are defined in figure 4 caption (symbols indicate means).

# Tables

**Table 1** Summary of metrics for the cohort and 3 example patients for the evaluative target volume (ETV), skin, ribs, chest, breast, heart, and lung.

| | | ETV | | | | Skin | | Ribs | Chest | Breast | Heart | Lung |
|---|---|---|---|---|---|---|---|---|---|---|---|---|
| | | $D_{90}$ | $V_{100}$ | $V_{150}$ | $V_{200}$ | $D_{1cm2}$ | $D_{max}$ | $D_{max}$ | $D_{max}$ | $V_{150}$ | $D_{1cm3}$ | $V_5$ |
| | | Gy | (frac) | (frac) | (frac) | (Gy) | (Gy) | (Gy) | (Gy) | (cm$^3$) | (Gy) | (frac) |
| Full cohort (35 patients) | **MCref** | 91.7 | 0.894 | 0.637 | 0.341 | 84.4 | 187.9 | 746.3 | 927 | 33.0 | 4.38 | 0.0724 |
| | **TG43sim** | 105.1 | 0.928 | 0.748 | 0.474 | 69.9 | 189.2 | 168.8 | 693 | 44.5 | 2.05 | 0.0396 |
| | %Δ$_{avg}$ | -14.1 | -4.02 | -18.3 | -40.1 | 22.2 | 13.54 | 82.3 | 29.6 | 36.0 | 70.6 | 44.8 |
| | %Δ$_{std}$ | 5.83 | 2.53 | 5.96 | 11.3 | 12.7 | 20.03 | 6.74 | 41.2 | 7.7 | 23.4 | 21.0 |
| | IQR(**MCref**) | 26.4 | 0.0957 | 0.194 | 0.124 | 46.1 | 189 | 541 | 718 | 18 | 5.64 | 0.0692 |
| | IQR(**TG43sim**) | 33.8 | 0.0872 | 0.189 | 0.191 | 48.1 | 213.3 | 129 | 1053 | 22.7 | 1.80 | 0.0331 |
| Example: patient 6 | **MCref** | 82.05 | 0.858 | 0.439 | 0.185 | 24.3 | 31.8 | 2293 | 2293 | 44.39 | 7.86 | 0.151 |
| | **TG43sim** | 88.95 | 0.894 | 0.542 | 0.274 | 15.6 | 21.6 | 1084 | 1228 | 60.38 | 1.65 | 0.068 |
| | %Δ | -8.41 | -4.2 | -23.6 | -47.8 | 36.0 | 32.2 | 52.72 | 46.44 | 36.02 | 79.03 | 55.3 |
| Example: patient 14 | **MCref** | 85.8 | 0.883 | 0.665 | 0.383 | 116.5 | 152.8 | 319.5 | 1204 | 63.24 | 7.20 | 0.0894 |
| | **TG43sim** | 96.6 | 0.924 | 0.742 | 0.525 | 91.1 | 121.5 | 55.43 | 1392 | 75.10 | 6.22 | 0.0687 |
| | %Δ | -12.5 | -4.6 | -11.7 | -37.3 | 21.8 | 20.5 | 82.7 | -15.6 | 18.7 | 13.6 | 23.2 |
| Example: patient 30 | **MCref** | 119.7 | 0.997 | 0.786 | 0.414 | 176.9 | 386.6 | 277.5 | 380.4 | 33.56 | 5.74 | 0.0671 |
| | **TG43sim** | 144.7 | 1.000 | 0.944 | 0.631 | 160.0 | 376.5 | 41.50 | 268.1 | 47.82 | 1.48 | 0.0270 |
| | %Δ | -20.9 | -0.305 | -20.0 | -52.6 | 9.6 | 2.62 | 85.0 | 29.5 | 42.5 | 74.2 | 59.7 |

Abbreviations: %Δ = percent difference between **MCref** and **TG43sim** (see Eq. (1)); %Δavg = average percent difference over the cohort; %Δstd = standard deviation of percent differences; IQR = interquartile range of absolute doses (difference between third and first quartiles); frac = fraction of volume